# The Validity Window of Space-Charge-Limited Current Measurements of Metal Halide Perovskite Devices

William R. Fisher[1*], Philip Calado[1], Jason A. Röhr[2], Joel A. Smith[3,4], Xingyuan Shi[1], Onkar Game[4], Jenny Nelson[1], Piers R.F. Barnes[1*].

[1]Department of Physics and Centre for Processable Electronics, Imperial College London, South Kensington Campus, London, SW7 2AZ, UK

[2]Department of Chemical and Biomolecular Engineering, Tandon School of Engineering, New York University, Brooklyn, New York 11201, USA

[3]Department of Physics, Clarendon Laboratory, University of Oxford, Oxford, OX1 3PU, UK

[4]Department of Physics and Astronomy, University of Sheffield, Sheffield, S3 7RH, UK

*Communicating authors: w.fisher17@imperial.ac.uk, piers.barnes@imperial.ac.uk

## ABSTRACT

Space-charge-limited current (SCLC) measurements are used to estimate charge carrier mobilities and electronic trap densities of semiconductors by analysing the current density-voltage (JV) relationship for unipolar devices predicted by the Mott-Gurney (MG) law. However, the interpretation of SCLC measurements for metal-halide perovskites is problematic due to mobile ionic defects which redistribute to screen electrostatic fields in devices during measurements. To overcome this, an SCLC measurement protocol was recently suggested that minimises ionic charge redistribution by probing the current during millisecond voltage pulses superimposed on a background bias. Here, we use drift-diffusion simulations with mobile ions to assess the validity of the MG law for analysing both the standard and new protocol JV measurements. We simulated idealised perovskite devices with differing mobile ion densities and compared them with simulations and measurements of devices with typical contact materials. We found the validity region for the MG law is limited to perovskites with mobile ion densities lower than the device's equilibrium charge carrier density (<$10^{17}$ cm$^{-3}$ for 400 nm thick methylammonium lead iodide films) and contacts with injection/extraction barriers ≤ 0.1 eV. The latter limitation can be partially overcome by increasing the device thickness, whereas the former limitation cannot. This restricts the range of perovskite layer compositions and viable contact materials that can be reliably analysed with the MG law. Approaches such as estimating trap densities from the apparent voltage onset to trap-free SCLC regime should also be critically reviewed since they rely on the same potentially invalid assumptions as the MG law. Our results demonstrate that extracting meaningful and accurate values for metal halide perovskite material properties from SCLC maybe challenging, or often not possible.

# I. INTRODUCTION

Metal halide perovskite materials are of considerable interest for applications including solar cells [1,2], photodetectors [3], light-emitting diodes [4], and memristors [5]. Given the wide variety of compositions possible amongst this material class, there is great interest in developing accessible methods to measure properties such as charge carrier mobility, charge carrier concentration and trap density as new perovskite variants are developed.

Superficially, one of the most common approaches to estimate these parameters is via space-charge-limited current (SCLC) measurements; analysis of current density *vs* voltage (JV) measurements of two-terminal perovskite single-carrier devices. Besides the numerous factors that complicate the analysis of SCLC measurements of non-perovskite devices, such as trap states [6,7], doping [8] and energetic disorder [9], the high density of mobile ionic charge in metal halide perovskites [10–12] further complicates the application of this measurement technique and the ensuing analysis. Adapting the SCLC measurement technique and/or analysis to account for these mobile ions to find electron and hole mobilities in perovskites would, therefore, be of great interest.

In principle, the JV curve of a semiconducting layer sandwiched between symmetrical and Ohmic contacts demonstrates SCLC beyond a voltage threshold, this SCLC region of the curve is known as the Mott-Gurney drift regime. JV curves are generally measured by stepping sequentially (sweeping) through an increasing or decreasing series of applied voltages, $V$, at some rate while recording the resulting current density between electrical contacts, $J(V)$. SCLC analysis usually involves determining the log-log gradient of the measured JV curve as it varies with voltage, $g(V) = \mathrm{d}log(J)/\mathrm{d}log(V)$, to identify the exponent of the power-law relationship $J \propto V^g$. Voltage regions where $g(V) \approx 2$ are identified and analysed using the Mott-Gurney (MG) law [13],

$$J = \frac{9}{8}\mu\epsilon_r\epsilon_0 \frac{V^2}{L^3} \qquad [1]$$

to extract a charge carrier mobility, $\mu$, assuming the thickness of the perovskite layer, $L$, and the relative and vacuum permittivities, $\epsilon_r$ and $\epsilon_0$, are known. Further modifications to this analytical framework account for the effects of electrode/semiconductor energy barriers, electronic trap states and regimes other than SCLC in the JV measurement [7,14–16]. For many devices, neglecting these contributions results in the incorrect interpretation of JV curves, namely identifying regions where $g(V) \approx 2$ as space-charge limited when they are not, and then applying the standard SCLC analysis (equation 1) to find $\mu$.

However, the application of the MG law and its associated analyses on metal halide perovskite materials is further confounded the presence of high concentrations of mobile ionic defects within the crystal lattice. These ions redistribute in response to varying internal electric fields during measurements [10,17], changing the electrostatic profile in the device, and hence the observed JV curve. Although there are examples of SCLC-type analyses and other associated techniques being applied to metal halide devices in the literature, the presence of mobile ionic charge is likely to invalidate the interpretation and conclusions drawn from the data [18–22]. Recent works [23–25] challenged the validity of the application of the MG law and associated analyses to perovskite devices and suggested experimental techniques to account for the effects of mobileions. Here we will investigate and review the arguments presented in these publications.

In this article, we show that applying the MG law to JV measurements of mixed electronic-ionic conducting materials such as metal halide perovskites is only valid under specific measurement conditions and when the electronic charge density is significantly greater than the mobile ionic charge density in the device at equilibrium (herein, 'equilibrium' is defined as the condition where no (net) energy or mass flows in or out of the defined system). For the typical class of materials and devices of interest to the perovskite community it is difficult to ensure these conditions are met, hence limiting the utility of the MG law and associated techniques in this context. We demonstrate these findings by presenting simulations of JV measurements of 'ideal' thin-film devices (which are not necessarily always possible to make) over a large voltage range (which is experimentally impractical) and by contrasting them to experimental results from real thin-film devices and corresponding simulations of them. This is done for two SCLC measurement techniques: (i) JV sweeps performed by incrementally stepping the applied voltage with time at a fixed rate and (ii) the so-called "pulsed-voltage" method where voltages are applied in very short pulses, probing the device in a quasi-fixed state. Noting the discrepancy between the ideal simulations and the experimental results, an analysis revealing the validity regime for the MG law is provided. The effects of increasing perovskite layer thickness on these validity limits are also investigated. The consequences of this validity regime and the machanisms at play have important implications for any JV measurement of perovskite single-carrier devices.

## II. RESULTS AND DISCUSSION
### A. Ionic Contribution to the Electric Field Distribution

The derivation of the Mott-Gurney law requires two assumptions: i) that the spatial variation of the internal electric field is governed only by injected majority electronic charge (the majority charge carrier here is holes) with no contribution from other charged species, and ii) that the electronic current is dominated by drift. While the first assumption should be valid in an ideal device at reasonably high voltages, a high density of mobile ionic charge, which might be expected in metal halide perovskites, is likely to invalidate this assumption since we expect the ionic charge to dominate the electric field profile in a device. The electric field profile, $E$, can be determined from Gauss' law which is related to the spatially varying distribution of charge. If the electronic charge in the perovskite is dominated by free holes of concentration, $p$, and the perovskite also contains a mobile cationic charge species with concentration, $c$, compensated for by an equally and oppositely charged, uniformly distributed static anionic species, with concentration $a$, then at steady-state:

$$\frac{dE(x)}{dx} = \frac{q}{\epsilon_r \epsilon_0}[p(x) + c(x) - a] \qquad [2]$$

Where *x* refers to the position along the one dimensional space of the perovskite layer and *q* is the elementary charge. The corresponding steady state drift current density, *J*, through the device will be given by:

$$J = q\mu_p p(x) E(x) \qquad [3]$$

where $\mu_p$ is the mobility of holes. Note that while ions contribute to the overall electric field driving the drift current, there is no direct contribution from mobile ions to the conductivity, i.e., $q\mu_p p$. We assume there are no other ionic interactions with the device interfaces (such as chemical reactions). Following the approach of *Schottky and Prim* [26], which analyses a case where no ionic charge is present, equations 2 and 3 would be combined by substitution and then integrated to derive the MG law. However, this method is no longer possible because the mobile ion density cannot be removed from the resulting equations or correctly accounted for without knowing its spatial distribution.

The MG law also assumes that JV measurements are made on semiconducting layers interfaced with Ohmic electrical contacts. Ohmic contacts maximise the concentration of electronic charge in the device at equilibrium and minimise any barriers to electronic charge injection, assuming all other variables (contact area, interfacial surface/dipole states etc.) are equal. We define an ideal Ohmic contact as one with no workfunction offset ($\phi_{offset} = 0\ eV$):

$$\phi_{offset} = |E_{BE} - \phi_m| \qquad [4]$$

where $E_{BE}$ is either $E_c$ or $E_v$, which are the conduction and valence band edge energies of the semiconductor, respectively, and $\phi_m$ is the workfunction of the metallic contact. The convention we use here is to discuss the smallest offset of the two according to the contact selectivity of carrier species in the unipolar devices. Variation of $\phi_{offset}$ impacts the charge carrier density in the device. In an out-of-thermal-equilibrium semiconductor with quasi-fermi levels $E_{Fn}$ and $E_{Fp}$ and with effective density of states $N_c$ and $N_v$ for electrons and holes respectively, the Blakemore approximation [27,28] gives:

$$n \approx N_c \frac{1}{\exp\left(-\frac{E_{Fn}-E_c}{k_B T}\right)+0.27} \quad \text{and} \quad p \approx N_v \frac{1}{\exp\left(-\frac{E_v-E_{Fp}}{k_B T}\right)+0.27} \qquad [5]$$

where $k_B T$ is the thermal energy. At the device interfaces $E_c - E_{Fn}$ or $E_{Fp} - E_v$ are equal to $\phi_{offset}$. The Blakemore approximation is used instead of the Boltzmann approximation in this calculation of the charge carrier density because for small $\phi_{offset}$ (of less than a few k$_B$T) Boltzmann statistics are no longer valid. The use of the Blakemore approximation is discussed in the **Methods** section and the entire approach of calculating the charge carrier density of semiconductor devices is discussed in the **supplementary material** where an upper-limit for charge carrier densities in perovskite devices is also estimated without such assumptions for the Blakemore distribution function. Throughout this work, one can view $\phi_{offset}$ as a proxy for majority charge carrier density with an inverse and exponential relationship. As we will show later in (Section II, part D), if the contacts to the perovskite are not perfectly Ohmic (i.e., when $\phi_{offset} > 0\ eV$) the MG law's window of validity in the presence of mobile ions is further reduced.

## B. JV Sweeps

To investigate the consequences of mobile ionic charge for the application of the MG law and its associated analyses we present drift-diffusion simulations (using our modelling package, Driftfusion [29]) of JV sweeps on thin-film metal/perovskite/metal devices. The contact energy levels were set such that holes were the majority charge carrier (see figure S1a-b), although the results are generalisable to electron-only devices. All device parameters and material constants for these ideal devices are presented in table SI. Figures 1a-c show results for the device depicted by the inset in figure 1a in which the 400 nm thick perovskite layer has mobile ions but no electronic trap states and has ideal Ohmic contacts with $\phi_{offset} = 0\ eV$ between the metal contacts and the perovskite valence band. A voltage range of 0 to 500 V is applied to investigate the full range of device behaviour, albeit noting that a 500 V applied to a real-world device of this thickness would casue damage and is not realistic in practice.

The example simulated JV curves for forward (from 0 to 500 V) and reverse (from 500 to 0 V) voltage sweeps at different scan rates in figure 1a show hysteresis caused by the mobile ionic charge. The lowest scan rate shows the least hysteresis, indicating that the ions have almost reached a quasi-static distribution for each voltage sampled independent of scan direction. At higher scan rates the ions have insufficient time to respond and thus screen the applied potential, resulting in more hysteresis. These curves also show a higher threshold voltage at which the gradient of the JV curve increases from its initial low-bias slope during the forward scan. The corresponding $g(V)$ values, as a function of forward scan voltage are shown in figure 1b (in the reverse scan $g(V) \approx 1$ for the fast scans and is similar to the forward scan at low scan rates). The regions of the g(V) curves that fall within the grey band in figure 1b are equivalent to $g(V) = 2 \pm 0.2$, this equates to $J \propto V^2$ where the MG law may be applied while allowing for some experimental error. We analysed these regions using the MG equation (equation 1), to obtain the hole mobilities that would be inferred from SCLC measurements (figure 1c). This range of $g(V)$ values is chosen puposefully to represent a reasonable range of gradients for which a researcher might consider applying the Mott-Gurney law to analyse the JV curve due to experimental errors or insufficient data points/range. For comparison, the hole mobility defined in the simulations (20 cm$^2$ V$^{-1}$ s$^{-1}$) is also shown (dashed red line). It is evident that mobility values inferred from the forward scans depend strongly on the scan rate, and that applying the MG law to a slow JV scan can overestimate the mobility by more than two orders of magnitude (~2000 cm$^2$ V$^{-1}$ s$^{-1}$) due to a different distribution of ionic charge corresponding to each voltage. We conclude that these gradient regions could easily be misinterpreted as falling within the MG regime and regions that lie above the grey-shaded area in figure 1b, i.e. $g(V) > 2$, could be misinterpreted as being due to charge-trapping effects [18,20,22]. At sufficiently high scan rates we approach the limit where ionic charge is unable to respond to the rapidly changing field in the device (maroon curves in figures 1b and 1c), which corresponds to ionic charge that is effectively frozen in its initial distribution during the JV scan. In this state, the extracted mobility is close to the mobility set in the simulation. This suggests that if a device could be measured with the ions effectively frozen in their equilibrium distribution, then the MG law and other associated analysis might be meaningfully applied. In the following section we will investigate this hypothesis.

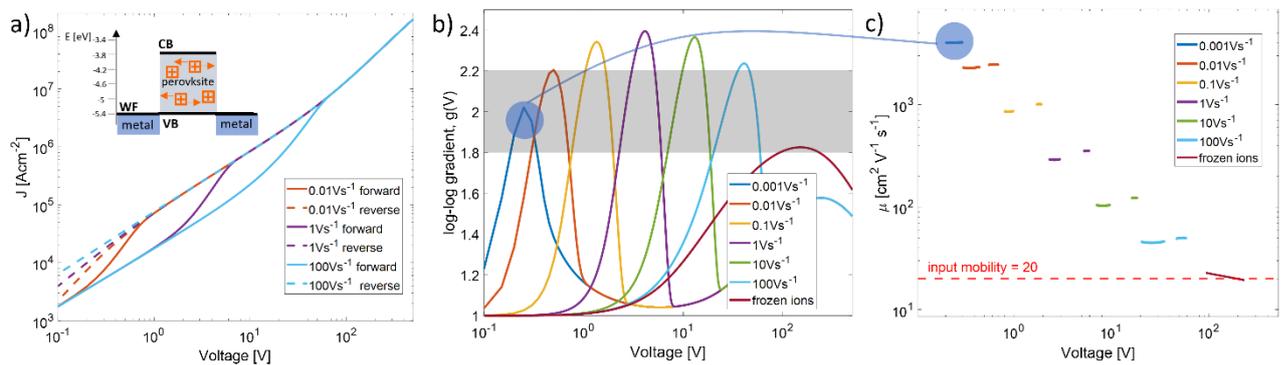

**Fig. 1**  Simulated JV sweeps and their analysis. (a) Example JV sweeps at three scan rates of a device represented in the inset with one mobile ionic species. (b) Calculated *g(V)* of the JV sweeps' forward scans in panel (a) with additional scan rates presented, the grey box represents a region of $g(V) = 2 \pm 0.2$ where one may attempt to apply the MG law to calculate $\mu$ from experimental data. (c) The corresponding $\mu$ calculated using the MG law from the regions of the JV curve where $g(V) \approx 2$ in the grey box of panel (b). The simulation parameters are listed in the ***supplementary material*** in table SI.

We now consider experimentally measured JV curves of perovskite hole only devices. Figures 2a-c display results for a thin-film device of architecture shown in the inset of figure 2a, with an approximate thickness of 400 nm for the perovskite layer. The chosen materials and processing techniques are those commonly used in relevant photovoltaic research, where additional thin selective transport layers are included to make the contacts closer to Ohmic than they would otherwise be. Details of the device fabrication are presented in the ***methods*** section. The maximum applied voltage was limited to 1.5 V to minimise bias-induced degradation. Figures 2d-f display the simulation results for a device approximately the same as the aforementioned experimental counterpart; the energetics and charge carrier densities from the simulation are shown in figures S1c-d. These results show an agreement of their general behaviours and trends with the experimental results in figures 2a-c, despite some apparent discrepancies. Concerning these trends, the simulations show the same dependence of JV hysteresis on scan rate, they show a large peak in the *g(V)* values and also show a similar estimation of the mobility, albeit all shifted to a lower voltage range. Meanwhile, there are differences between the simulations in figures 2d-f and the experimental results in figures 2a-c: the simulations give a slightly larger range of current densities and estimated mobilities over the JV sweep and the location of the *g(V)* peaks is at a lower voltage in the simulations. These differences are likely due to the omission of charge trapping and interfacial states in the simulations, which will be present in real devices, and the error on the values assigned to certain physical input parameters. The estimated mobilites are $\mu_p \approx 10^{-3}\ cm^2V^{-1}s^{-1}$ from the experiments (figure 2c) and $\mu_p \approx 10^{-2}\ cm^2V^{-1}s^{-1}$ from the simulations (figure 2f), we not the values are imprecise due to the large variance in *g(V)* throughout the scans. These values are orders of magnitude smaller than values of electronic carrier mobilities typically reported for MAPbI$_3$ films in the literature ($\mu \approx 7.5\ cm^2V^{-1}s^{-1}$ [30] or $\mu = 10 - 44\ cm^2V^{-1}s^{-1}$ [31]). A recent communication by Xia *et al.* addressing the mobilities of lead-halide perovskites summarises reported electronic charge carrier mobilities in the literature, where the range of carrier mobility values for metal-halide perovskites given was between $\mu = 0.7 - 600\ cm^2V^{-1}s^{-1}$ [32]. Assuming the film morphology in this work is similar to those in the studies mentioned, this shows that our calculated mobilities substantially underestimate the actual mobilities. The general agreement between the full device simulations with the experiment implies that some of the non-ideal features of the experimental results (when compared to the results in figure 1) are likely due to the addition of selective transport layers and possibly due to $\phi_{offset}$ asymmetry, since these are the main differences between the simulated devices in figures 1a-c and 2d-f.

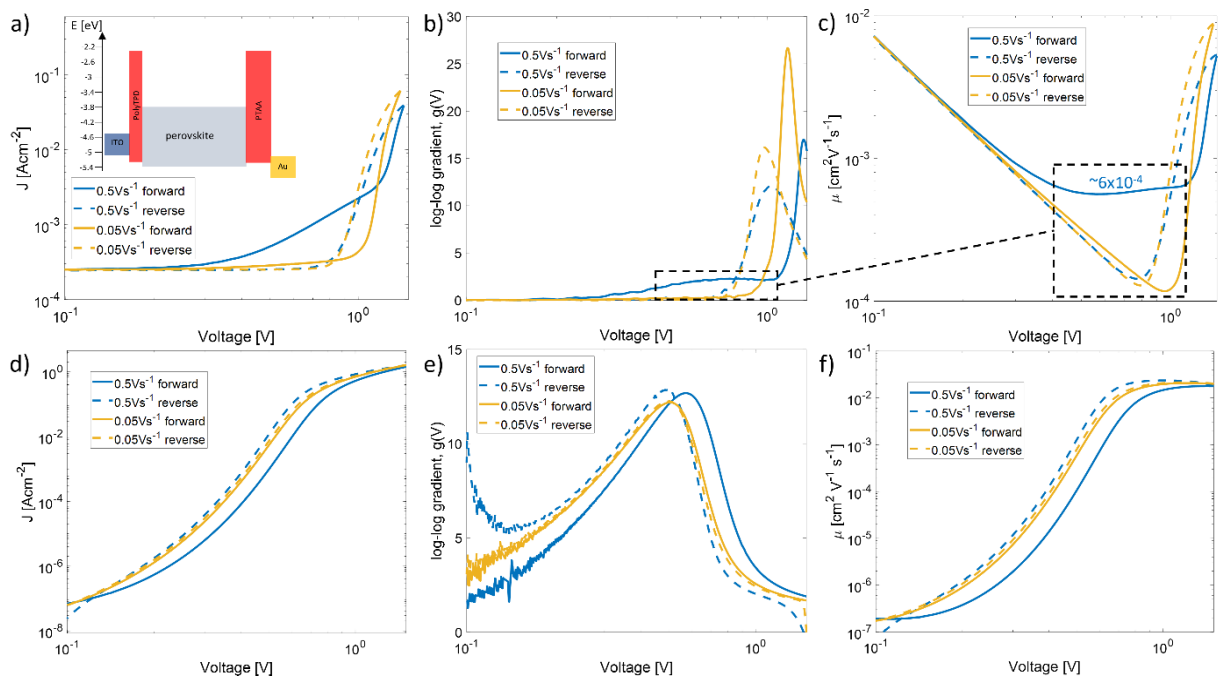

**Fig 2.** (a) JV sweep experimental results at two different scan rates of a device represented in the inset. (b) *g(V)* of the sweeps in panel (a). (c) $\mu$ calculated using the MG law and values from the sweeps in panel (a) with the region of interest highlighted by a black-dashed box. The region of interest is a guide to the eye and does not have the same meaning as the grey region in figure 1b. (d-f) are the simulation results of a device representative of the device investigated in (a-c) using the same analysis

techniques. The simulation parameters are listed in the ***supplementary material*** in table SII.

The ionic drift during the JV scans in figure 2 results in hysteresis and contributes to the substantial increases or decreases in current density at intermediate applied voltages. The changes in current due to ionic drift can be explained by the accumulation/depletion of ionic charge increasing or decreasing the electronic current flowing across the interface(s) with the direction and magnitude of this depending on the relative signs of the ionic and electronic charge carriers. This modulation of the electronic current by the accumulation/depletion of ions at the interface(s) might prevent SCLC conditions from arising within this chosen scan range. Additionally, the presence of hysteresis in these JV scans indicates the ionic charge is greatly influencing the electronic carriers' spatial distribution (and thus, the internal electric field and the electronic current density), invalidating assumptions of the MG law and contributing to $g(V) \gg 2$. It is the redistribution of ionic charge, not the injected electronic charge, primarily determines the electric-field profile in the device. We note that extended-range JV curves (figures S2a and S2b) display asymmetry between positive and negative applied voltages, which is influenced (partially) by the energetics of the contacts being asymmetric. The asymmetry of the JV shown in figure S2a has been shown by Game *et al.* [33] to also be a function of ionic drift during the measurement. Figure S2b is measured using the pulsed voltage technique (described in Section C.) which removes the contribtion from drifting ionic charge during the measurement. The results still show asymmetry, although less than is shown in figure S2a, suggesting that contact asymmetry exsists in this device and does play a role in the device behaviour along with the mobile ionic charge. We, therefore, conclude that the large peak in the *g(V)* and the underestimated mobility values are due to some combination of asymmetric contacts, the ionic drift during the scans and the additional selective charge transport layers, the exact contributions of these factors are difficult to decouple. We doubt the underestimated mobility is due to the limited scan range used in the experiments/simulations due to the significant plateau in estimated $\mu$ at voltages $> 1\,V$ in figure 2f and the significant decrease in $g(V)$ at voltages $> 1\,V$ in figure 2b. From figure 2 it is clear that JV curves of non-ideal single carrier devices give results that should not be analysed using the MG law and, if they are analysed in this way, it is likely that a large under-estimate of the charge carrier mobility is obtained.

Comparing the experimental and simulation results in figure 2 to the simulation of ideal devices in figure 1, reveals behaviour/trends in common: the direction of hysteresis in the JV curves is the same in both, and the location of the *g(V)* maximum in the forward scan is at a higher voltage when the scan rate is higher in both. We note that there are several differences between the results in figures 1 and 2: *g(V)* reaches very high values in the measured devices which was not reproduced in the simulations of the idealised device; the voltage at which the forward scan *g(V)* peaks is less scan-rate dependent than in the simulation of the ideal device, whilst there is a significant disparity in the magnitude of the current densities between the simulations and the experiments. These differences between figures 1a and 2a (when compared over a similar voltage range) particularly the latter, can be attributed to the existence of asymmetric contacts and selective transport layers. The limitations of available materials and processing techniques can clearly be seen.

Given the strong effect from mobile ions during the JV measurements and simulations, on both in the magnitude and shape of the JV curves, and the differences in JV hysteresis at different scan rates, it is clear that more sophisticated ways of measuring SCLC in perovskites must be employed. In the following section we investigate a recently developed pulsed voltage SCLC method and the limits of its applicability. An in-depth explanation of the technique is provided in the **methods** section.

### C. Pulsed Voltage Method

Recently, *Duijnstee et al.* published an approach for measuring JV curves without significantly altering the distribution of ions during the scan [24]. This technique involves applying a series of short voltage pulses typically of 10s to 100s ms, with each pulse separated by a return to a reference dc bias voltage (usually 0 V) for a long period of time (> 300 s). The short pulse durations aim to ensure that the ion distribution does not have time to deviate from the reference distribution set at each dc measurement voltage. This method is adopted in this section and referred to as the "pulsed-voltage method".

Figures 3a-c display simulation results where the ionic charge is frozen in its equilibrium (i.e., no (net) energy or mass flowing in or out of the defined system) distribution during the scan (see the **methods** section for a detailed explanation of the meaning of the equilibrium condition) but is mobile during the equilibration process. In figure 3a, we only present the forward scans because the reverse scans are identical due to the lack of ion drift during the simulated JV scan. These results are significantly different to those presented in figure 1, showing that the pulsed voltage method is significantly supressing the effect the ionic charge has on the JV curves. We do, however, see changes in the JV curves by varying the ion density: the magnitude of the currents observed, and $g(V)$ are affected by the concentration of ionic charge. The extracted MG mobility (figure 3c) is also affected, although the values obtained for this ideal device are within $\pm 50\%$ of the set mobility (i.e., 18-26 $cm^2 V^{-1} s^{-1}$ as compared to 20 $cm^2 V^{-1} s^{-1}$ as the numerical input). The effect of a static equilibrium ion distribution on SCLC characterisation is best demonstrated in figure 3b - the devices containing higher ion densities hardly reach a $g(V) = 2 \pm 0.2$ within the voltage range investigated. These results indicate that devices with high ion densities may not be suitable for the application of the MG law, since their JV curves are significantly different to those with no ionic carriers, and they never reach the same $g(V) \approx 2$ as the case with no ionic carriers, even if the technique can produce reasonable estimates for the mobility.

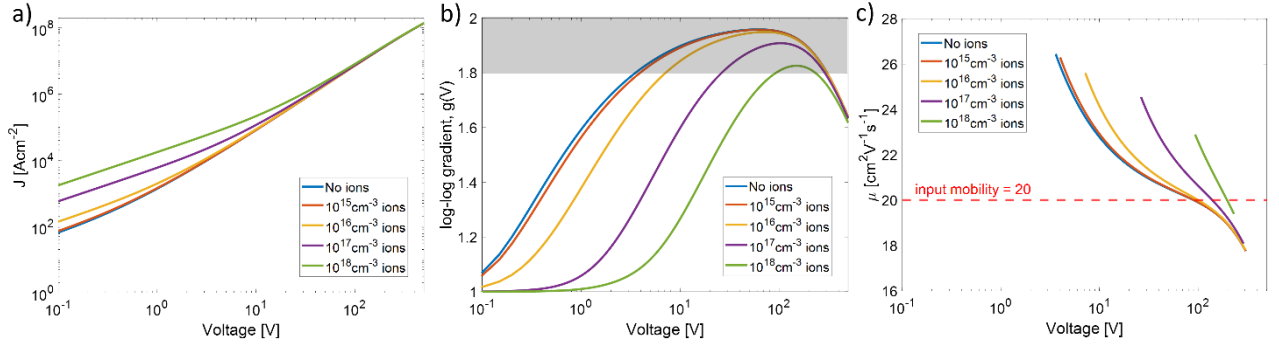

**Fig.3** Drift-diffusion simulations of JV curves with the mobile ionic charge frozen in its equilibrium position. (a) Simulated JV curves of the same device as in figure 1a with varying ion concentrations, we present only the forward scans because the reverse scans are identical due to the lack of ion motion. (b) Calculated *g(V)* of the JV curves' in panel (a). (c) $\mu$ calculated using the MG law from the regions of $g(V) \approx 2$ in panel (b).

In figure 3a, as the input ion density increase, the current density increases at low voltages. The mobile (during equilibration) ionic charge allows for more electronic charge to exist in this single carrier device at (and near) equilibrium, the details of which will be addressed in future work. Figure 3b shows that in the intermediate voltage regime (up to $\approx 100\ V$) the current-limiting processes are the same as in a device with no ions; however, because of the ionic charge, such space-charge limited current either occurs at different voltage onsets or is even obscured. At higher voltages ($> 100\ V$) in our simulations we observe that a combination of the charge-extraction coefficient at the electrodes (representing the interfacial resistance of a metallic electrode) and the electronic carrier saturation limit [16] starts to limit the increase in current. Both effects could occur in an experiment if these high voltages could be achieved. For the higher ion densities at very high voltages this current limitation dominates the response before any SCLC effect (fully) emerges, as the latter are obscured by other processes. This supports the proposition that the MG law and other associated analyses are not valid for perovskites with high ion densities: this is even in the absence of electronic traps, injection barriers and with ions frozen in their equilibrium state.

The equilibrium distribution of the ionic charge has a significant effect on the simulated JV curves. As electronic charge redistributes to align the Fermi level in the perovskite layer with the contact workfunctions at the interfaces, the ionic charge redistributes too—it contributes significantly to the space-charge layers at the interface(s) in the same way as electronic charge; either by accumulating or depleting. This not only reduces the density of electronic charge at the interfaces when compared to a semiconductor with no mobile ionic charge, but also modifies the energetic landscape at the interfaces by locally increasing the charge density, reducing the space-charge layer thickness and, in turn, modifying the electric field in the space-charge region. The simulated electric field and electrostatic potential profiles of devices at equilibrium and at 75 V during a JV scan with frozen ions are shown in figures S3 and S4 respectively. These show significant contribution of the ionic charge to the potential distribution and electric field distributions of the bulk and at the interface, even when the ions are spatially fixed during the simulated JV scan.

Figures 4a-c show experimental results for the case where ionic charge is nominally frozen during the JV scan using the pulsed voltage technique. These measurements were carried out on the same device as in figure 2 and the data were obtained by a pulsed voltage method which is illustrated in the inset of panel 4a. The results from this pulsed voltage method are substantially different to those of the JV sweeps shown in figures 2a-c: the ionic charge being 'frozen' has greatly reduced hysteresis (it may be that the small yet discernable hysteresis seen in figure 4a is caused by the ionic charge not being entirely stationary for the short pulses and/or some experimental artefacts), yet we still see $g(V) > 2$. In fact, $g(V)$ extracted from the experimental data (as in figure 4b) is somewhat erratic and never consistently $\approx 2$. For simplicity, we calculate the mobility over the entire voltage range of 0.2-1.4 V, as shown in figure 4c. The calculated estimates of hole mobility are $\sim 2 - 5 \times 10^{-4}$ cm$^2$V$^{-1}$s$^{-1}$, sitting orders of magnitudes lower than those reported in the literature, as with the mobilities determined from the JV sweeps (figure 2). This suggests that despite the alternative measurement technique (using pulsed voltages) this does not really make MG analysis applicable or effective for this class of thin-film devices, since the calculated mobility is wrong by orders of magnitude. We present reasons for this in the following.

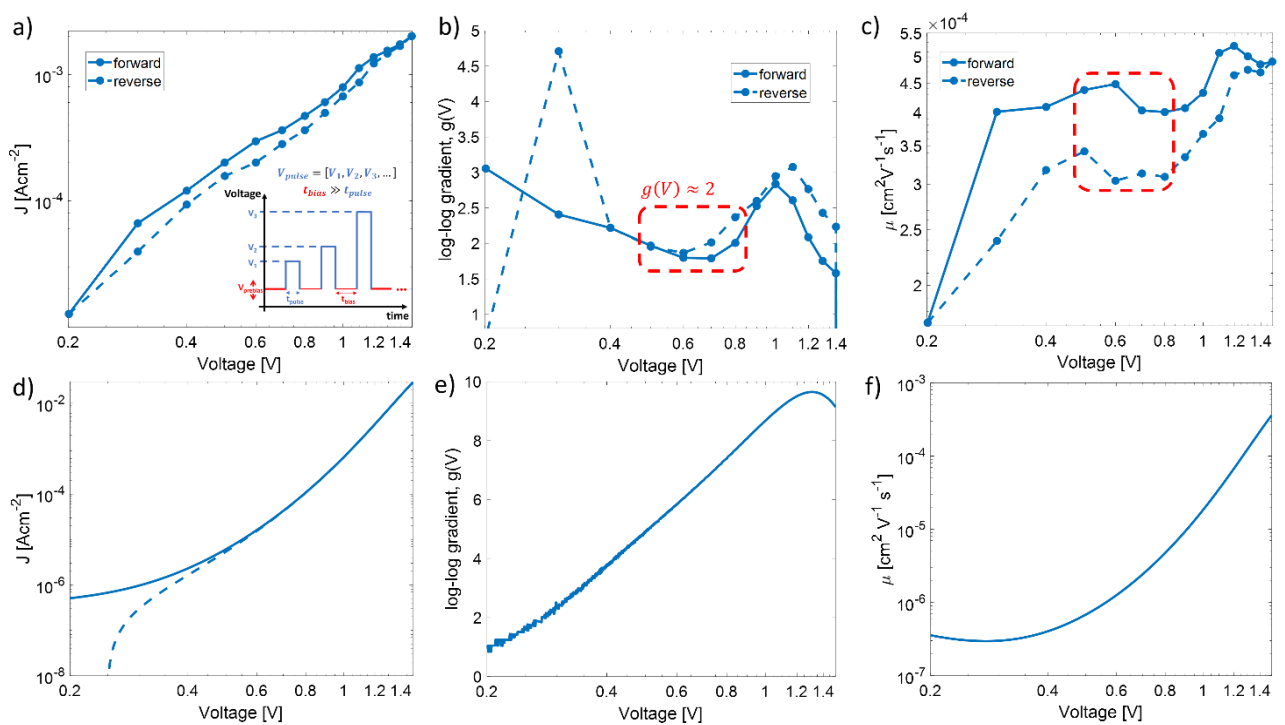

**Fig.4** (a) Experimental JV results of the same device represented in figure 2a but using the pulsed voltage method depicted in the inset. Both forward (solid) and reverse (dashed) scans are shown. (b) *g(V)* of the curves in panel (a). (c) *μ* calculated using the MG law and values from the curves in panel (a). (d-f) are the simulation results of a device representative of the device investigated in panels (a-c) using the same analysis techniques with an input $\mu = 20\ cm^2\ V^{-1}\ s^{-1}$. The divergence between the forward and reverse scans at low voltages in panel (d) corresponds to capacitive charging/discharging with the frozen ions and is only visible at these very low total current densities. Corresponding simulation parameters are listed in the *supplementary material* in table SII.

Figures 4d-f show simulations of the experimental case presented in figures 4a-c. The disparity in current density between simulation and experiment that is present in figure 2 is also seen here for the same reasons. Even though through most of the voltage range we see similar current densities, estimated mobilities and *g(V)* peak locations (1.1 - 1.2 V), the simulation shows a much higher *g(V)* at the peak (∼9.5) than that of the experiments (∼3), indicating that either the parameters of the simulation are not entirely representative of the real device or that the ionic charge is not completely frozen in the experiments, as it is in the simulations. The input mobility (20 $cm^2 V^{-1} s^{-1}$) and the mobilities (figure 4f) derived from the *JV* curve (figure 4d) are still orders of magnitude apart, showing that the MG law is ineffective at estimating the mobility of the majority charge carrier in this device and confirming the findings of the experimental results in figures 4a-c. In order to demonstrate that ionic charge, asymmetric contacts and the selective charge transport layers are still significantly affecting these results we carry out the same sets of simulations as in figures 4d-f while varying the ion density and then as well as removing the selective charge transport layers. These results are shown in figures S5a-c and S6a-c respectively. With this further investigation we find that in the full devices the ionic charge density in the perovskite layer changes the size and location of the peak of *g(V)* (the effects of the built-in potential), reduces the current density and reduces the estimated mobility, even though the ionic charge is frozen during the JV scan. We also find that removing the selective transport layers only slightly increases the current density and estimated mobilities: the energetics of the metal contacts still matter and, therefore, the reasons for the non-ideal experimental results are indeed the combination of the ionic charge, the additional selective transport layers and the asymmetric contacts.

Altogether, the simulation and experimental results in figure 4 further confirms that thin-film devices of this type are not suitable for analysis with the MG law and even though using the pulsed voltage measurement technique makes a significant difference to the JV curve, it is not sufficient for allowing the MG law to be applicable. We conclude that ionic charge, even when frozen in its initial position in an ideal device, still has a large impact on the JV measurement of the device and the utility of the application of the MG law to it: the simulations show that as mobile ion densities increase, the low voltage Ohmic current is increased and the quadratic power law behaviour at high voltages is suppressed. We note that the effects of uniform, static doping on these devices, as described by Röhr and MacKenzie [8], show the same trends and are similar to, but not the same as, the effects due to ions seen here. In cases where conventional electrical doping is present in the perovskite [34] in addition to the mobile ions explored here, the effects of each would occur simultaneously with the strongest of the two dominating the device behaviour. The experimental results suggest that thin-film devices and/or other devices without ideal, symmetrical, ohmic contacts (and especially with aditional selective transport layers) should not be characterised with the MG law, even if the ionic charge is static during the measurement by using the pulsed voltage method. We further propose that the original analyses derived from the MG law that use the same set of assumptions should not be used for analysing, for example, the so-called "trap-filled limit" [7].

While experimental results of the pulsed voltage method on a thin-film single-carrier device did not yield correct mobilities, the SCLC analysis of simulated JV curves for ideal devices with lower concentrations of 'frozen' mobile ions (figures 3a) yielded mobilities close to what was expected (the input mobilty). We will now investigate the validity region for SCLC measurements for which the application of the MG law could meaningfully be applied in the frozen ion regime.

### D. Is there a region of validity for SCLC measurements of metal halide perovskites?

In order to determine a region of validity for using the MG law for analysing SCLC measurements, a parameter search is conducted: a series of simulations of the same ideal device shown previously where the contact workfunctions are varied symmetrically, i.e., the workfunctions of both contacts are always equal and while $\phi_{offset}$ is increased, and the ionic density is varied. As before, JV measurements from 0 to 500 V with the ionic charge frozen in their equilibrium distribution are simulated. The maximum $g(V)$ in the JV is found (depending on the simulated device this peak could occur anywhere in the range) and the voltage and current density at which this maximum occurs are used to calculate the mobility using the MG law. For our simulation results, this process selects for the calculated mobility in the JV curve closest to the set mobility (20 cm$^2$ V$^{-1}$ s$^{-1}$). If a region of the JV curve does not fall within $J \propto V^{2 \pm 0.2}$, i.e., $1.8 < g(V) < 2.2$ (as depicted in figure 5a)), then this is contained in the red shaded box depicting a region falling outside $J \propto V^2$ experimentally. The values of $g(V)$ for this set of simulations are given in figure S7. The results are presented as a heat-map in figure 5a. Considering that these devices are ideal and should give close to perfect mobility values, these results show that a good estimate of the mobility can be obtained from the MG law for regions where $g(V) \approx 2$ for a device with no traps or interfacial states if the contacts are very close to perfectly Ohmic and where the ionic charge carrier density $\leq 10^{18}$ cm$^{-3}$. We also find that ideally the ion density should be $\ll 10^{18}$ cm$^{-3}$, otherwise the region where $1.8 < g(V) < 2.2$ is very small and the contacts must be perfectly Ohmic. There is a trade-off between how perfectly Ohmic the contacts must be (the amount of $\phi_{offset}$) and the ion density: if the $\phi_{offset} > 0.08$ eV, the ion density must be $< 5 \times 10^{15}$ cm$^{-3}$ for MG analysis to remain valid. We note that the exact values of the above quantities depend on the assigned density of states and effective mass of electrons and/or holes (which are values commonly used in the literature) but the trends will still persist even if the density of states changes significantly. If the device falls outside of these given limits, the JV curve will not reach the SCLC regime since $g(V)$ never reaches the range $1.8 < g(V) < 2.2$ (even if the ions are frozen) and these devices cannot be meaningfully analysed using the MG law or other related analytical frameworks.

We also extended investigations into studying thicker devices (1 μm and 100 μm) undergoing JV scans up to 1500 V (shown in figures S8a and S9a-c respectively) revealing that the constraints on $\phi_{offset}$ can be stretched by having a thicker device; the region of validity in figure S8a is stretched in the y-axis when compared to figure 5a. Figures S9a-c investigate a device with $\phi_{offset} = 0.3\ eV$ and show no region of the curves with $1.8 < g(V) < 2.2$ (figure S8b) but the extracted mobilities at the maximum of $g(V)$ (figure S8c) yield very accurate results, we note that the application of the MG law is not valid in these regions due to the lack of SCLC, even though the mobility estimate is accurate. Figure S10 investigates devices of $\phi_{offset} = 0.2\ eV$ with varying thicknesses, confirming the observations from figures S8 and S9 that the large offset can yield SCLC regions but with very minimal changes to the validity region with respect to ion density (figure S10b). Figure S10 also confirms the observation from figure S9c that, curiously, the MG law can yield accurate $\mu$ estimates at $g(V)$ maxima (figure S10a) without any clear sign of SCLC, the cause of this observation is yet to be understood. Figure S10c does, however, reveal that very thick devices ($> 10\mu m$) may not be realistically measureable due to the very high voltage required to reach the maximum value of $g(V)$ in the JV curves.

Based on the observation that the region of validity for SCLC measurements depends on $\phi_{offset}$ and the ion density, and by noting that $\phi_{offset}$ is related to electronic charge carrier density (equation 5), we conclude that the relative density of the ionic to electronic charge must be substantial. A refined limit of these relative densities is given below.

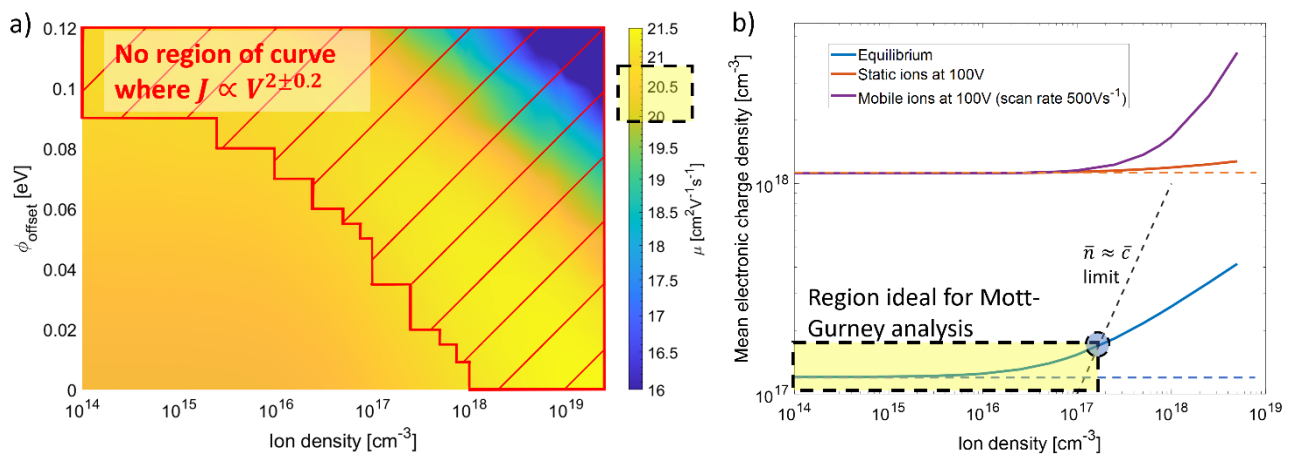

**Fig. 5** Analysis of drift-diffusion simulations showing the limits of the application of MG analysis to perovskite devices of 400 nm thickness. (a) A heat map representing the accuracy of the calculated mobility using the MG law for a set of contact energies (workfunction offset) and a set of ion densities. The set mobility is 20 cm²V⁻¹s⁻¹. The region outside the colour map is represented in red shading. (b) Calculated mean electronic charge density for perovskite layers of different ion density at two points during a JV scan (0 V and 100 V). The black box with yellow shading represents a region where the current density and the internal electric field of the device is defined by electronic charge. This is true because the ion density does not have an appreciable effect on the electronic charge in the device at 0 V and 100 V. By examining the ion densities covered by the black box in (b) and noting the region of the same ion densities in figure (a) we can see an area of yellow/orange, indicating the value of mobility is the same as the input mobility (20 cm² V⁻¹ s⁻¹). In this region of yellow/orange, we can see a set of parameters that gives the mobility calculated using the MG law that yields the correct result. This region is the part of the colour bar in yellow/orange highlighted by the black box. The boxes in panels a) and b) are showing cause and effect. This is a parameter set ideal for the application of the MG law. The blue circle in panel (b) shows the intersection of the calculated charge density at equilibrium with our suggested limit for MG law analysis: $\bar{c} < \bar{n}$ (black dashed line).

The explanation of the observation of ion density limiting $g(V)$ in a device even when it is static during the JV scan is shown in figure 5b. Figure 5b presents the average electronic charge carrier density ($\bar{n}/q$, where $q$ is the electronic charge) in the perovskite layer of 400 nm thickness and is calculated using:

$$\bar{n} = \frac{q}{L}\int_0^L [n(x) + p(x)]\, dx \qquad [6]$$

The results are for points at 0 V and 100 V (to represent an arbitrary point along the JV curve) during the same set of JV simulations (results at 20 V are given in figure S11a). First we focus on the red and blue curves where the ion density is frozen. The subtraction of these curves is equal to the charge injected into the device during the JV scan (results showing the amount of injected charge explicitly are shown in figure S11b and S11c). As the ion density increases in the perovskite, the electronic charge density at 0 V increases; the same happens to the electronic charge density at 100 V (and 20 V) until we reach the approximate limit where the ion density is comparable in size to the electronic charge carrier density at equilibrium ($\approx 10^{17}\ cm^{-3}$ for both at 0 V). This limit is depicted by the vertical, dashed black line in figure 5b. At this limit the amount of injected charge starts to decrease as the high density of static ions at the interfaces limits charge injection: the subtraction of the red and blue traces decreases past this limit. This explains the apparent trade-off between workfunction offset and ion density in figure 5a. The electronic charge injection limitation by the ionic charge (figures S3 and S4 show the cause) begins at an even lower ion density for devices with higher $\phi_{offset}$ because there is less electronic charge in the device at equilibrium. Upon analysing the magnitude of injected charge (shown in figures S11a and S11c) we note that the influence of the ionic charge on the injected charge begins before $\bar{n} \approx \bar{c}$, this allows a more refined limit to be placed. We, therefore, propose a theoretical guide/limit for MG analysis: the electronic charge density initially (at 0 V) in the perovskite must be approximately 5 times higher than the ionic charge density for the analysis to be applicable in the frozen ion measurement regime ($5\bar{c} < \bar{n}$). This guide/limit is likely to be conservative, but it ensures that the limit is generally true and independent of the density of states for electrons/holes.

To demonstrate the implications of this claim, we can estimate an upper limit to the electron charge carrier density in a perovskite device with ideal Ohmic contacts at equilibrium and compare this to likely concentrations of ionic charge in perovskites. We take the density of free electrons expected when the Fermi level reaches the perovskite conduction band edge as this upper limit for the average concentration at equilibrium ($n_0$). Details of this estimation are shown in the **supplementary material**. We obtain an approximate upper limit of $n_0 \leq 2 \times 10^{18}\ cm^{-3}$ for MAPbI$_3$. We note that the effective mass of holes in these materials is similar to or higher than that of electrons (0.29$m_e$ [35]) which would give a higher upper limit for the equivalent hole device (a factor of ~1.5 times higher). As stated above, ion densities must be approximately 5 times lower than the device's equilibrium electronic carrier concentration for the Mott-Gurney law to apply to an ideal device in the frozen ion regime, this gives an upper limit on ion density of $\bar{c} < 4 \times 10^{17}\ cm^{-3}$. The ion density would have to be lower than this for devices without ideal Ohmic contacts, since the 0 V electronic charge carrier density would be lower. To further demonstrate the $5\bar{c} < \bar{n}$ limit, figure S8b shows that in cases of non-ideal Ohmic contacts, a thicker device and low ionic charge density, the electronic charge density is significantly higher than the ionic density where the MG law could be meaningfully applied.

The purple trace in figure 5b shows the electronic charge density for a JV scan where the ions are mobile. The scan rate is selected such that hysteresis at 100 V for high ion density devices is very large and is very high but is necessary for demonstrating these effects when applying such high voltages, this also shows that very fast scan rates do not necessarily achieve the same results as the pulsed voltage technique (red trace). Similar discrepancies between the red and purple traces can be found at much lower voltages with much lower scan rates (see figure 1a), this example demonstrates one of the ways in which the electronic charge density is greatly influenced by mobile ionic charge near the voltages analysed in figure 5a. In this example a very large amount of electronic charge injection is induced by the redistributing ionic charge. At other points in the JV scan (or at different scan rates) the magnitude of this influence changes and results in the hysteresis we see in figure 1a. This result, further to the results from figure 1, shows that the amount of injected electronic charge is very dependent on ion drift in devices with higher ion densities and that SCLC analysis should not be performed without frozen ions. The purple trace in figure 5b also shows that if the ion density in the 400 nm perovskite layer is very low ($\leq 10^{15}\ cm^{-3}$ in this case) compared to the electronic charge density (by many orders of magnitude), there is little difference between a JV sweep and the pulsed voltage method, so a fast JV scan is an appropriate measurement technique.

We now comment on the constraints implied by our limit $5\bar{c} < \bar{n}$. Such Ohmic contacts would be very hard to find because of MAPbI$_3$'s $E_c \approx -3.8\ eV$ and $E_v \approx -5.4\ eV$. Examples of feasible metallic contacts for thin film devices are platinum and calcium. None of the metals commonly used for solar cell research are viable. As demonstrated earlier in the text, the constraints on $\phi_{offset}$ can be stretched by increasing device thickness and, therefore, gold and aluminium (giving $\phi_{offset} \approx 0.3\ eV$) may work for very thick devices. Unfortunately, the thicker the device, the higher the voltage required to reach the MG regime (this is the example shown in figure S9 and evident in figure S10c ).

The density of mobile ionic charge in metal halide perovskites is quoted as high as $10^{18} - 10^{19}\ cm^{-3}$ in some theoretical models and experimental results [10–12,36–40] but in some other studies as low $\approx 10^{16} - 10^{17} cm^{-3}$ [41–43] and sometimes even lower ($\approx 10^{13}\ cm^{-3}$) [25]. Determining mobile ion densities accurately is still challenging [44]. If this range of values is true, then it suggests that the validity of the MG law, which we have shown requires mobile ion density to be less than approximately 5 times equilibrium electronic density, may depend on layer morphology and perovskite composition. In many cases, though, it will not be known whether this mobile ion density limit is satisfied.

Finally, we note that the estimated mobilities in large portions of the red shaded box of figure 5a are reasonable and even though these device simulations produced no regions of the JV curve where $g(V) \approx 2$, the mobility determined from the peak of the JV is still very close to the value entered in the simulations (20 cm$^2$V$^{-1}$s$^{-1}$). The only region where the determined mobility significantly deviates from the correct value is where $g(V)$ approaches 1.5 (these regions are shown in figure S5). This observation, along with the estimated mobilities in the very thick devices of figure S10a show that there is more research to be done to fully understand the details of these results.

### III. CONCLUSIONS

We conclude that even measurement techniques that do not allow the ionic charge to move and that only probe electronic responses to applied voltage do not guarantee the applicability of analytical SCLC theory to the results. In its equilibrium position, increasing the mobile ionic charge increases the amount of electronic charge in the device but decreases the amount of injected charge which results in the MG regime not being reached and/or the wrong mobilities being calculated. The ionic charge accumulates at the interfaces as part of the space-charge regions that are normally formed at equilibrium, increasing the space-charge density and decreasing the width of the space-charge regions. This alters the local electric field in the space-charge regions. We have identified the controlling factor impacting the MG law's applicability as the relative densities of the ionic and electronic charge in the device, namely the ionic charge density needs to be at least five times lower than the electronic charge density at equilibrium. If this condition is met by achieving sufficiently Ohmic contacts on a device with low ion densities (< 5 times electronic charge density), the ionic charge has little effect on any JV measurement so there is no need to modify the standard measurement procedure (with pulsed voltages, for example). The limits on contact energetics can be stretched by increasing perovskite layer thickness. Unfortunately, very thick devices ($>> 1\mu m$) require applied voltages with which it is experimentally inaccessible to reach the MG regime. Other features in real devices, including trap statesand interfacial states, can affect SCLC measurements. The precise nature and origin of electronic trap states in metal halide perovskites are still unclear and varies significantly between perovskite compositions; thus trap states are currently not well accounted for when modelling perovskite devices and analysing measurements of them. It is also challenging to find electrode materials that lead to ideal Ohmic contacts and the density of ionic charge (particularly for polycrystalline thin films) is difficult to control and likely to be too high. We, therefore, suggest that it is difficult to make a single-carrier thin-folm  perovskite device that can be analysed using the MG law and its associated analyses. We demonstrated this through our own set of experimental results and recommend that metal halide perovskite devices should only be analysed with the MG law and associated techniques when the mobile ion density and contact energetics (workfunction offset) fit within the limits outlined by this work.

# Methods

## Pulsed Voltage JV Measurement

The measurement scheme is indicated in the inset of figure 2a which shows the device held at a pre-bias voltage, $V_{pre\text{-}bias}$, for a long time (in this work $V_{pre-bias} = 0$). Subsequently, a very short voltage pulse is applied, during which the current is measured, before returning to $V_{pre\text{-}bias}$. This process is repeated for a series of voltage pulses. The pulse duration ($t_{pulse}$) was set to 500 µs and the pre-bias time ($t_{bias}$) to 600 s. To extract the current density due to each voltage pulse, a linear fit to the current density transient for $50\ \mu s < t_{probe} < 250\ \mu s$ was performed and, by extrapolation, a value for the instantaneous current density at $t_{pulse} = 0\ s$ was found (see Figure S12). The series of probe voltages and the associated current densities are then plotted to create a JV curve.

## Drift-Diffusion Simulations

Device simulations are carried out using the open-source simulation code Driftfusion. This code allows for the simulation of semiconductors with electronic and ionic conduction. Details of the code, reasoning for the choices made in its construction and its capabilities can be found in Calado *et al.* [29].

*Finding the device's equilibrium condition* – This process is outlined in Calado *et al.* [29] but is included here for clarity. The device starts with a set of initial conditions and runs through several steps to find the equilibrium solutions. The equilibrium solutions take the form of the potential energy profile and the charge carrier densities as a function of position given no flow of energy or mass in or out of the defined system. For the ideal devices simulated in this work a linearly varying electrostatic potential, exponentially varying electronic carrier densities and a uniform density of anions and cations over the layer thickness are used for the initial conditions. These conditions ensure the boundary conditions are satisfied.

For the entire device (multi-layered) simulations (in the figures 2d-f and figures 4d-f) the initial conditions are different in that the electronic carrier densities are set to the equilibrium densities of the individual layers (defined by the set fermi level/doping density). This ensures the boundary conditions are satisfied. The equilibrium solution process is the same as for single layer devices.

*Distribution function* – A common distribution function used for drift-diffusion modelling of semiconducting devices is the Boltzmann approximation, which approximates Fermi-Dirac statistics, allowing for solutions for the charge carrier densities to be more easily obtained. This approximation is only valid when difference between the charge carrier's energy and the Fermi level is $\gg k_B T$. In some of the devices in this work this is no longer the case, so the distribution function is changed to the Blakemore approximation [27,28] including the terms for diffusion enhancement and setting $\gamma = 0.27$.

## Device Fabrication

**ITO/PolyTPD/PFN-P2/MAPbI$_3$/PTAA/Au/SiO$_2$**

All materials were purchased from Sigma Aldrich unless otherwise stated and used as received. ITO substrates were etched using Zn and 4 M HCl, and cleaned by sonication in dilute Hellmanex with boiling DI water for 15 minutes, rinsed with boiling DI water, sonicated in DI water, rinsed with acetone and sonicated in IPA. Substrates were dried with N$_2$ and UV ozone cleaned for 15 minutes. All fabrication was in a N$_2$ glovebox with continuous purging. PolyTPD (Ossila) solution (1.5 mg ml$^{-1}$ in dichlorobenzene) was spin-coated at 6000 rpm for 30 s (acceleration 2000 rpm s$^{-1}$) followed by annealing at 100 °C for 10 minutes. Next PFN-P2 (1-Material, 0.5 mg ml$^{-1}$ in methanol) was coated at 5000 rpm for 20 s (dynamic dispense). Acetonitrile:methylamine MAPbI$_3$ solution was prepared following the procedure by Noel *et al.* [45], with methylamine gas flowed through a 0.5 M solution with a 1:1 molar ratio of MAI (Greatcell) and PbI$_2$ (TCI) until dissolved. Films were spin-coated at 2000 rpm and annealed at 100 °C for 15 minutes. PTAA solution (10 mg ml$^{-1}$ in toluene) was prepared with additives (per ml) of 4 µl of tBP (4-*tert*-butylpyridine) and 7.5 µl of LiTFSI (from a 170 mg ml$^{-1}$ stock solution in acetonitrile) and stirred overnight. The PTAA was then spin-coated at 4000 rpm (dynamic dispense). Devices were completed by thermally evaporating 80 nm of Au (CooksonGold) at 0.1 -1.2 Å s$^{-1}$. Finally, SiO$_2$ (K J Lesker) encapsulation was deposited by e-beam evaporation at 1 Å s$^{-1}$ with a base pressure of 3E-6 mbar. A 150 nm layer was deposited in two 75 nm steps to minimise substrate heating.

## Acknowledgements

We thank Dr Davide Moia for his help with optimising circuit elements during the assembly of the experimental apparatus. We acknowledge funding from the UK Engineering and Physical Sciences Research Council (grants EP/M025020/1 , EP/R020574/1, EP/T028513/1, EP/R023581/1, EP/M014797/1).

## Supplementary Material

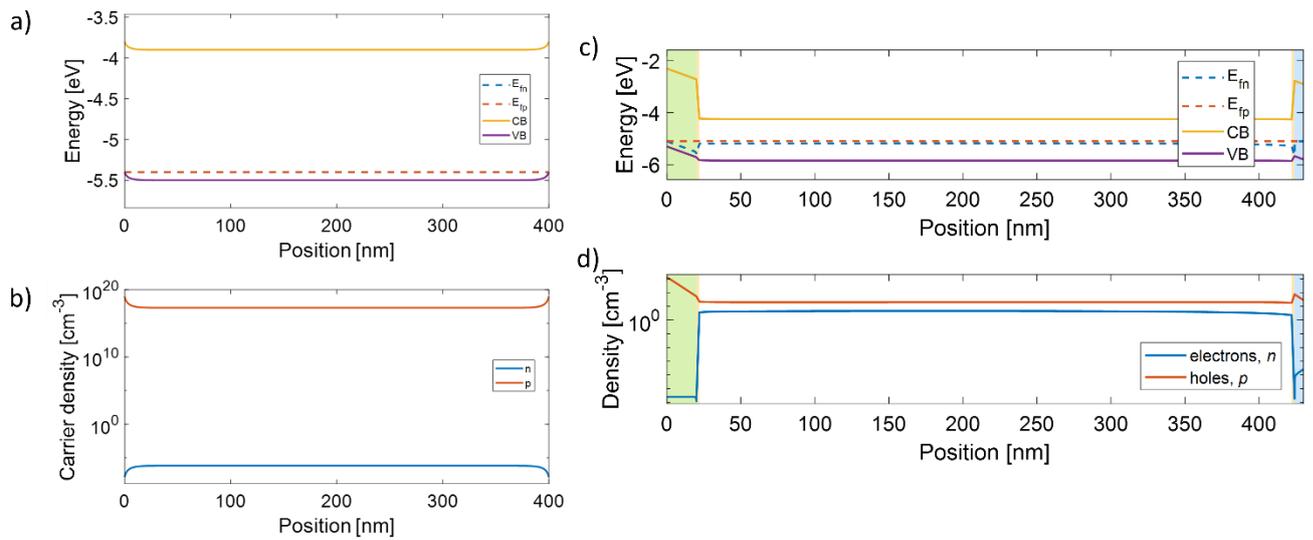

**Fig. S1a-b** Simulation results for the equilibrium solution of a device with symmetrical contacts with no workfunction offset at the valence-band (at -5.4eV) showing a) the energy levels and quasi-Fermi levels of the device and b) the charge carrier densities.

**Fig. S1c-d** Simulation results for the equilibrium solution of the device investigated experimentally in this work a) the energy levels and quasi-Fermi levels of the device and b) the charge carrier densities.

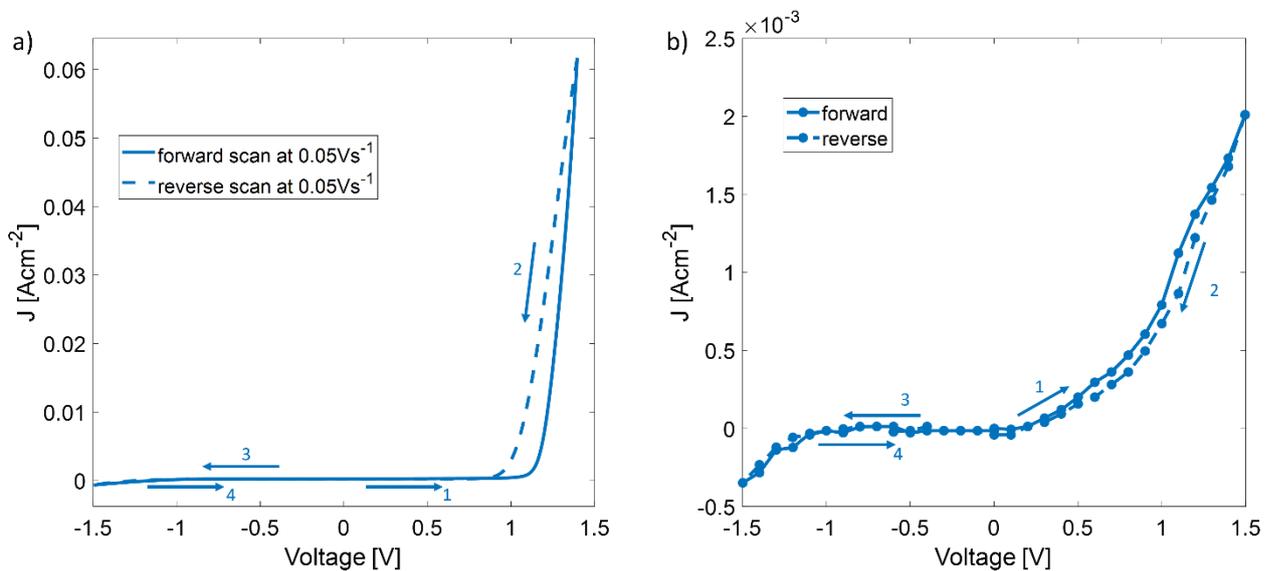

**Fig. S2** The experimentally determined JV curves from -1.5V to 1.5V of the thin-film device measured in figures 2 and 4. a) The JV sweep measured at $0.05\ Vs^{-1}$ in the order depicted by the numbered arrows starting from 0 V. b) the JV curve measured buy the pulsed-voltage technique in the order depicted by the numbered arrows starting from 0V. The asymmetry in the current density magnitude in both figures between $V > 0$ and $V < 0$ suggests that the device energetics are asymmetric.

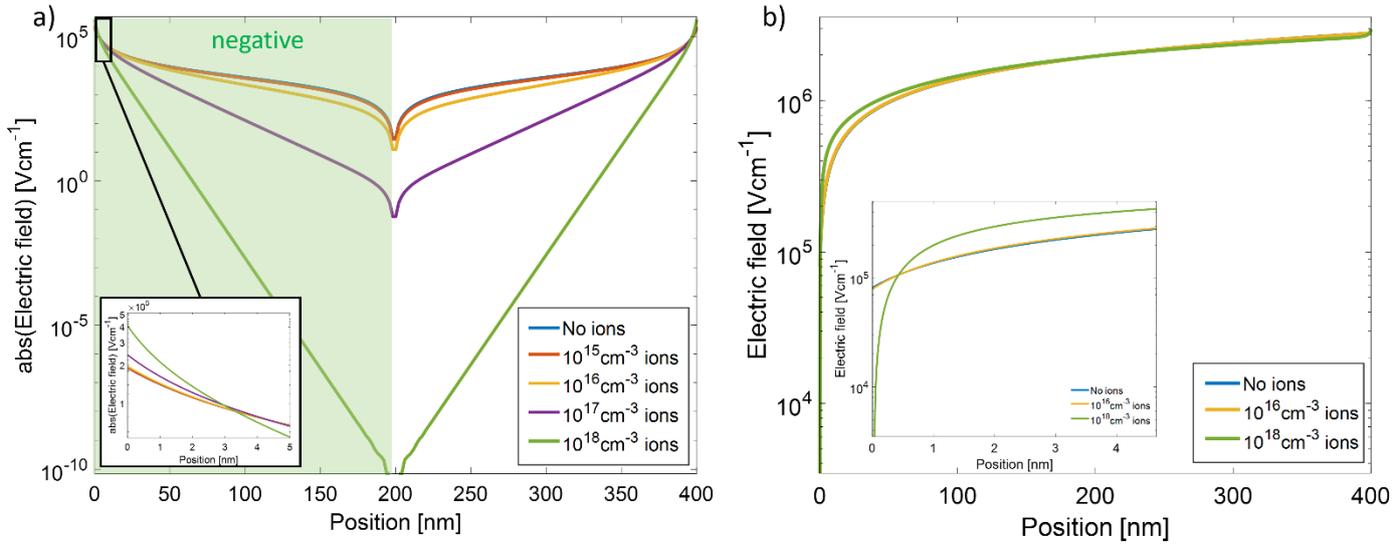

**Fig. S3** The Electric field profile in the ideal device simulated in figure 3. a) the magnitude of the electric field is shown, allowing for a logarithmic scale to be used. The deice is equilibrium here. b) at 75V during a JV scan with the ionic charge frozen in its equilibrium distribution. The insets of both figures show a "zoom in" on the region close to the left-hand interface.

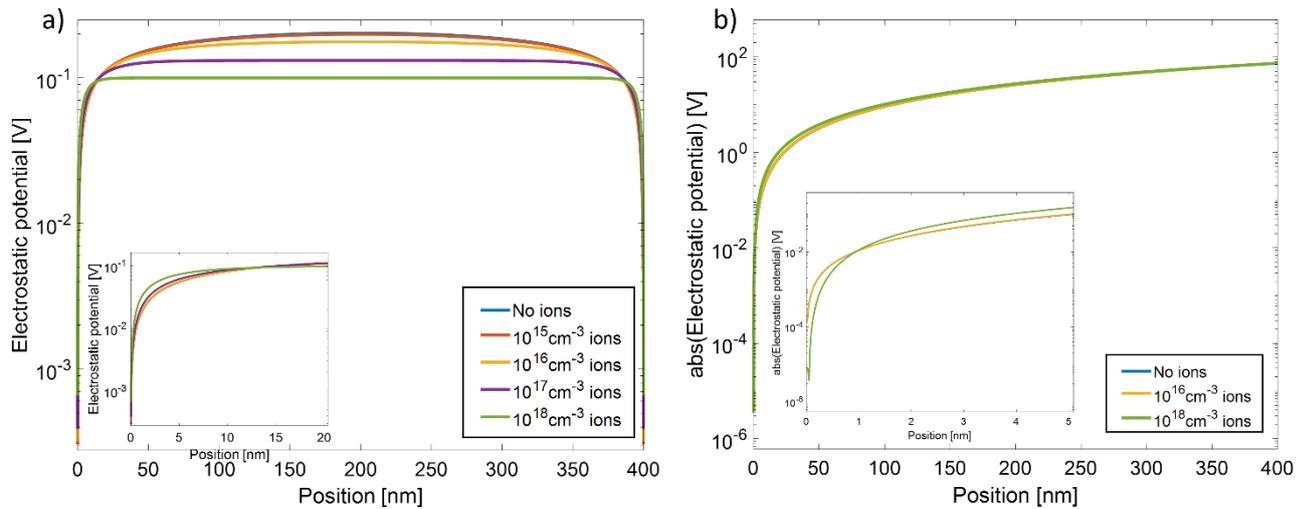

**Fig. S4** The Electrostatic potential profile in the ideal device simulated in figure 3. a) Electrostatic potential at equilibrium. b) the magnitude of the electrostatic potential is shown, allowing for a logarithmic scale to be used. Electrostatic potential is negative in for the whole device. The device is at 75V during a JV scan with the ionic charge frozen in its equilibrium distribution. The insets of both figures show a "zoom in" on the region close to the left-hand interface.

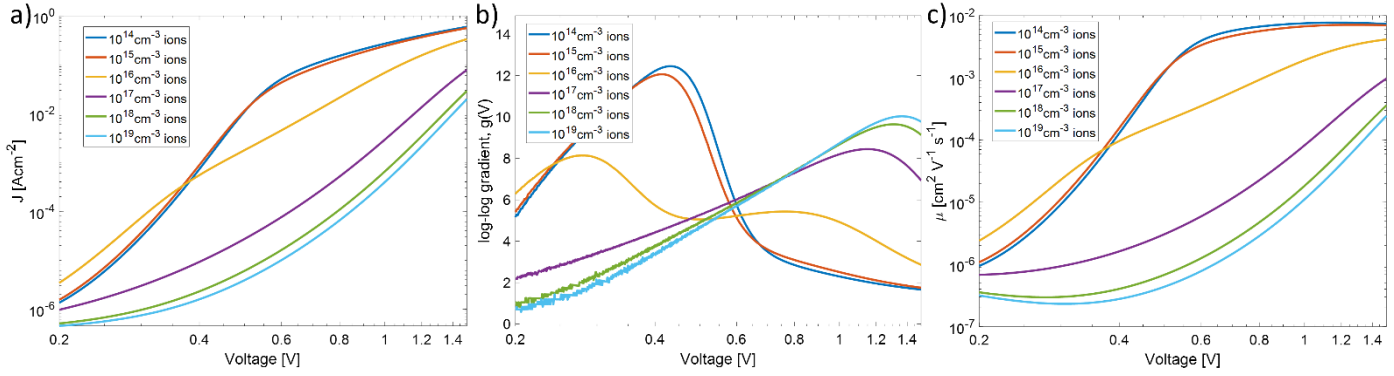

**Fig. S5** Simulations on the device used to produce the experimental results in this work with the parameters displayed in SII but varying ion density with the ionic charge frozen in their equilibrium position. (a) shows the JV curves, (b) shows the $g(V)$ values and (c) shows the mobilities calculated using the MG law. A heavy dependence of the results on ion density is shown.

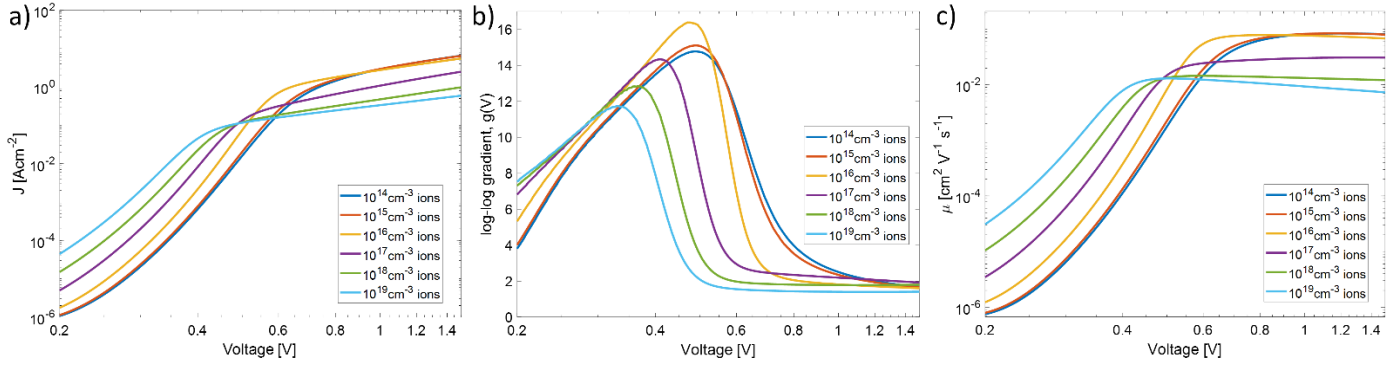

**Fig. S6** Identical simulations to those in figure S5 except the selective charge transport regions have been removed. (a) shows the JV curves, (b) shows the $g(V)$ values and (c) shows the mobilities calculated using the MG law. The results are still dependent on ion density but the behaviour of the higher ion density scans is very different to those in figure S5 particularly in the low voltage region where the high ion density cases have much lower current densities and $g(V)$ values than their equivalent simulations in figure S5.

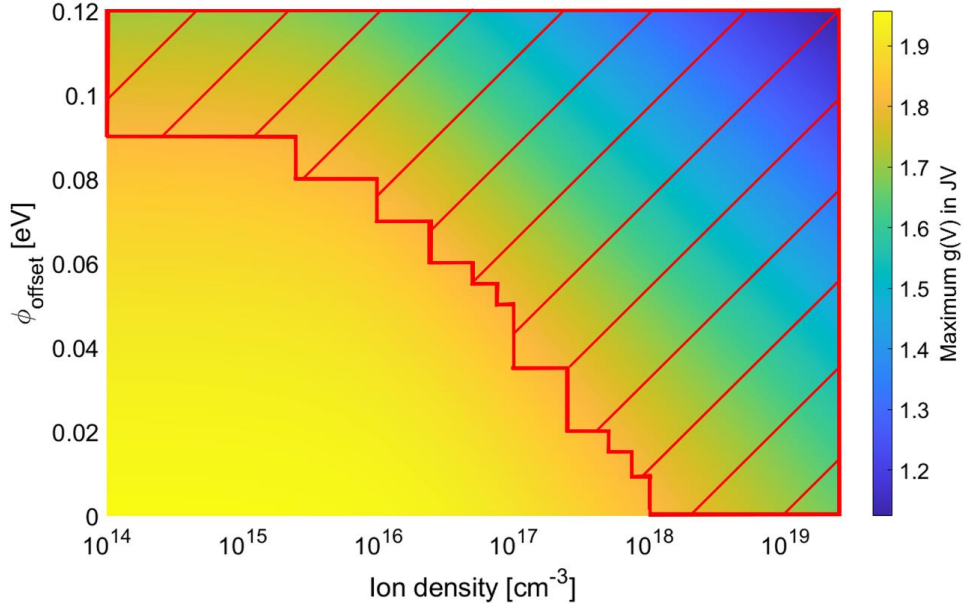

**Fig. S7** Analysis of a set of drift-diffusion simulations explaining the limits of the application of MG analysis to perovskite devices of thickness 400 nm showing the highest value of *g(V)* for each simulated JV curve for this set of devices. The red shaded box shows the region where $g(V) < 1.8$.

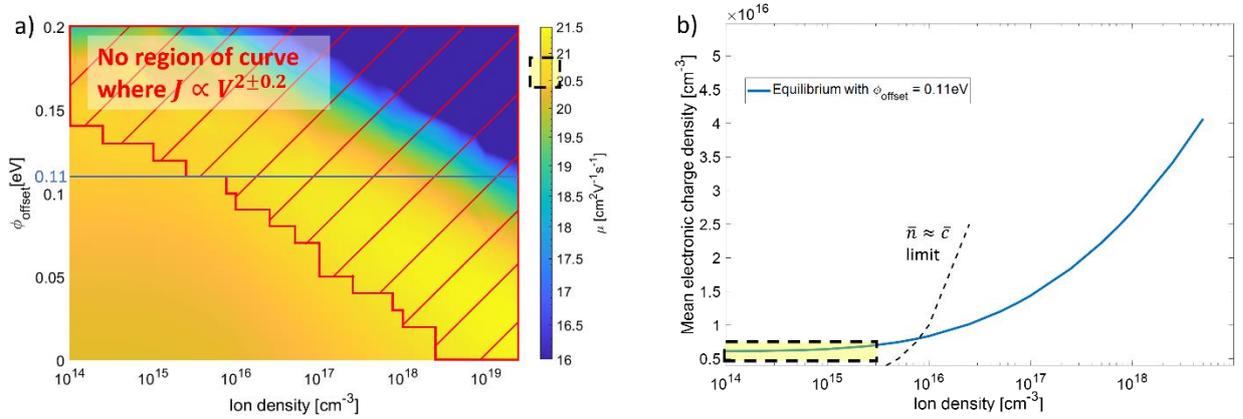

**Fig. S8** Analysis of a set of drift-diffusion simulations explaining the limits of the application of the MG law to perovskite devices of thickness $1\mu m$. (a) A heat map representing the accuracy of the calculated mobility using the MG law for a set of contact energies (workfunction offset) and a set of ion densities. The set mobility is 20 cm²V⁻¹s⁻¹. The red shaded region represents the simulated JV curves where *g(V)* never entered the selection zone for MG analysis. (b) The calculated mean electronic charge density for perovskite layers of different ion density at equilibrium at 0.11eV $\phi_{offset}$. The blue line on figure (a) is where the calculations to produce figure (b) are made. The region highlighted by a black-dashed box filled in yellow in (b) falls under the limit $5\bar{c} < \bar{n}$ and corresponds to the regions of the colour-map with mobilites withing the black-dashed box on the colour -scale in (a), this is where the MG law can be meaningfully applied.

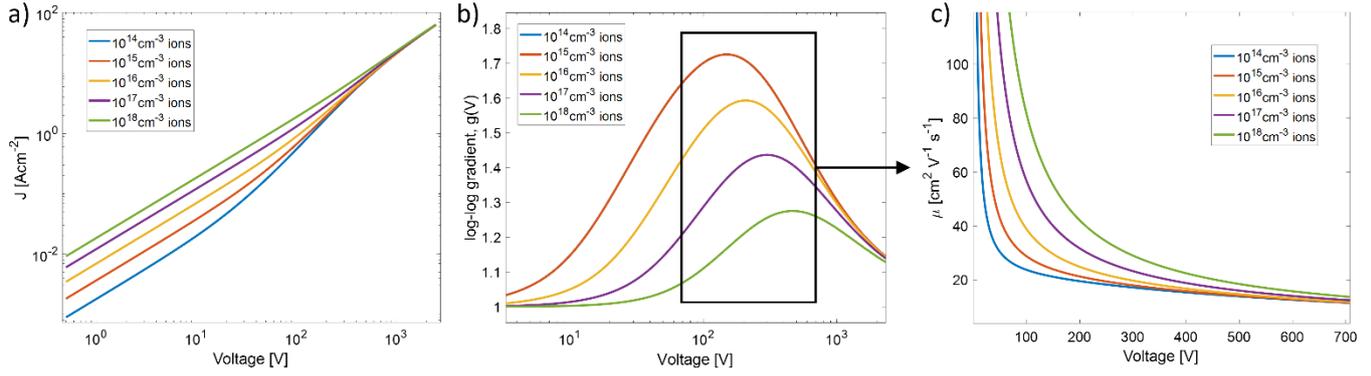

**Fig. S9** (a) JV simulations of a device with $\phi_{offset} = 0.3\ eV$ and thickness $100\mu m$ where ion density is varied. (b) the *g(V)* values calculated from the JV curves in (a) – note that the blue curve is very close to the red curve and is obscured. (c) the mobility values calculated from the region in the black box of (b): the region where the peaks in *g(V)* occur.

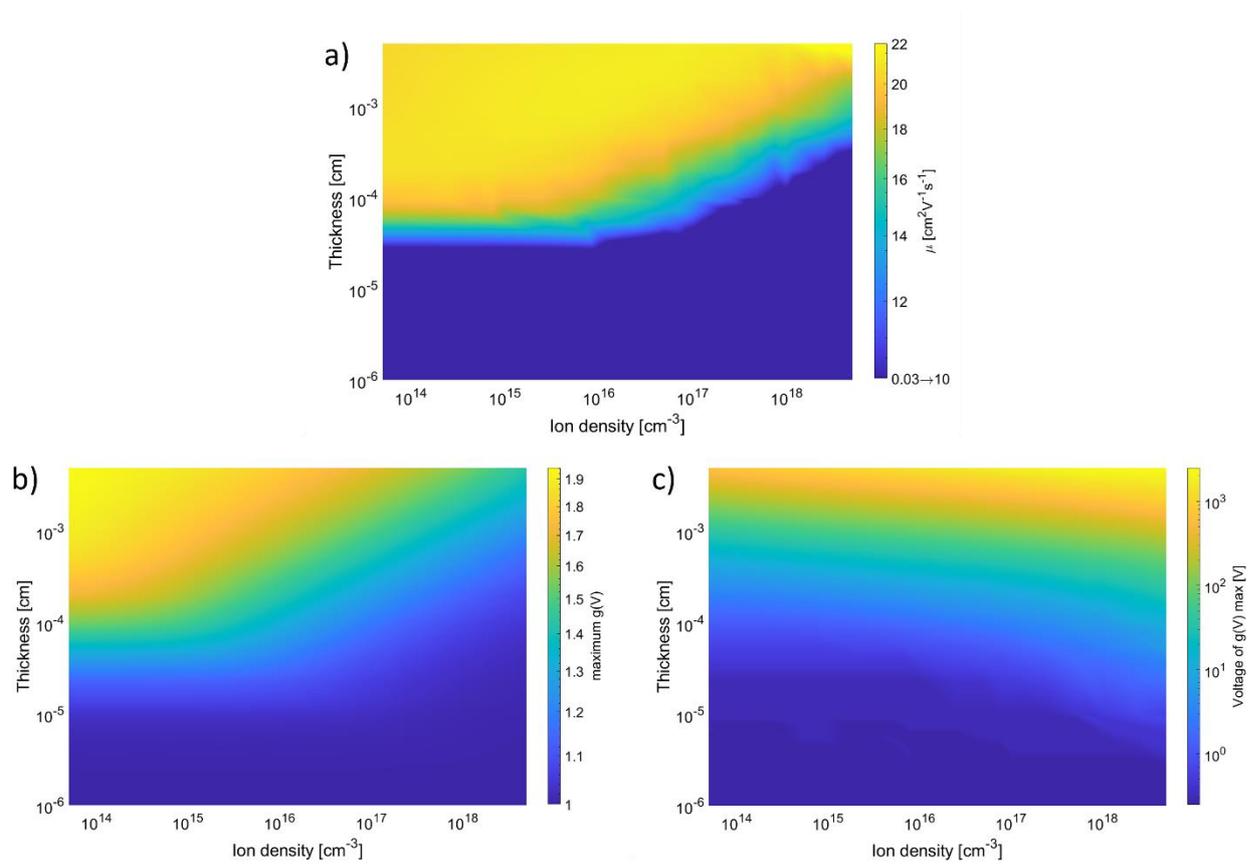

**Fig. S10** Heat maps of results from JV simulations to 2000 V with ionic charge frozen in their equilibrium distribution of a device with $\phi_{offset} = 0.2\ eV$ whilst varying the thickness of the active layer. (a) The mobility calculated using the MG law at the position of maximum g(V) in the JV curve. Note that the dark blue colour refers to a range of values from $\mu = 0.03 \rightarrow 10\ cm^2V^{-1}s^{-1}$, this was done such that the colour change is maximised near the mobility set in the simulations ($\mu = 20 cm^2V^{-1}s^{-1}$). (b) the maximum value of $g(V)$ during the simulated JV curves, the values of $g(V) > 1.8$ indicate a SCLC regime has been reached. (c) the voltage at which the maximum value of $g(V)$ occurred during the JV scan.

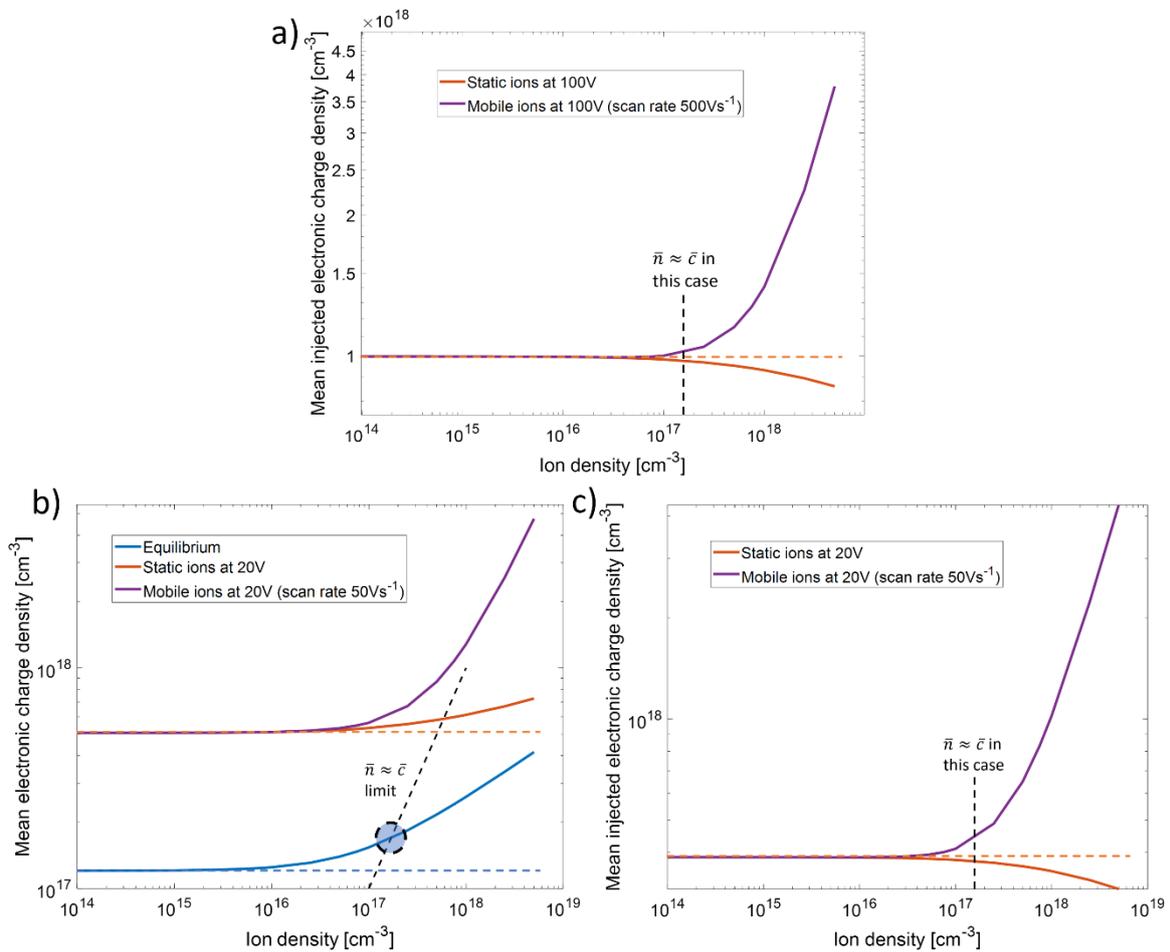

**Fig. S11** (a) the injected electronic charge averaged across the device thickness at 100V during JV scans *vs* ion density. The specific limit for MG analysis of this device is shown by the black dashed line. (b) the mean total electronic charge density at equilibrium and at 20V during JV scans, the blue circle represents the intersection of our suggested limit (black dashed line) and the charge density at equilibrium, for ion densities lower than at the blue circle, MG analysis will be valid. (c) the injected electronic charge averaged across the device thickness at 100 V during JV scans *vs* ion density. The specific limit for MG analysis of this device is shown by the black dashed line.

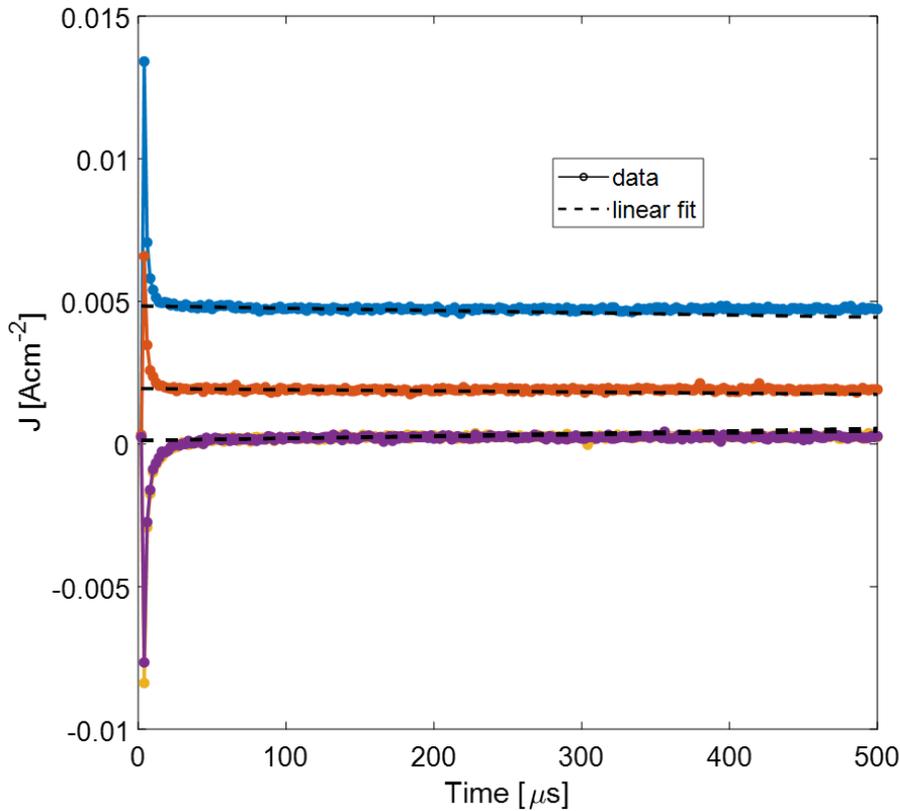

**Fig S12.** An example of current transients at four separate probe voltages (two negative and two positive). The dotted lines represent the linear fits to the data at times immediately after the capacitive response. The instantaneous current density is then inferred from the linear fit at *t* = 0 s.

## Calculation for the estimation of the upper limit of charge carrier densities in single carrier perovskite devices:

Derivations of this form are very common in many semiconductor textbooks and lecture series, the derivation used here is some arbitrary combination of many of them. In a device with ideal ohmic contacts the work function energies of the contacts are aligned with the conduction band (electron affinity) or valence band (ionization potential) energies of the semiconductor. Under these circumstances the maximum density of electrons or holes in the semiconductor can be estimated by finding the density when the Fermi level ($E_F$) is at the conduction or valence band edge. The electron density in a semiconducting material can be calculated from the integral:

$$n_0 = \int_{E_c}^{\infty} g(E)F(E)dE \qquad [7]$$

where $E_c$ is the energy of the conduction band edge, $g(E)$ is the density of states and $F(E)$ is the occupancy factor as a function of energy $E$. The equation for approximating the density of states near the conduction band edge of a semiconductor is also well known if the system is 3 dimensional and the dispersion relation is assumed to be parabolic:

$$g(E) = \frac{(m^*)^{\frac{3}{2}}\sqrt{2(E-E_c)}}{\pi^2 \hbar^3} \qquad [8]$$

where $m^*$ is the effective mass of the electron and $\hbar$ is the reduced Planck's constant. The occupancy factor is given by the Fermi-Dirac distribution (since Boltzmann statistics do not apply for $(E - E_F) \leq k_B T$),

$$F(E) = \frac{1}{1+e^{\frac{(E-E_F)}{k_B T}}} \qquad [9]$$

where $k_B T$ is the thermal energy. Equations [8] and [9] are substituted into [7] assuming $E_F = E_C$ and rearranged to obtain an integral known as the Fermi-Dirac integral of order ½ (here $F_{1/2}(\eta_F)$) giving,

$$n_0 = \frac{\sqrt{2}(m^* k_B T)^{\frac{3}{2}}}{\pi^2 \hbar^3} \int_0^{\infty} \frac{\eta^{\frac{1}{2}}}{1+\exp(\eta-\eta_F)} d\eta \qquad [10]$$

where the substitutions $\eta = (E - E_c)/k_B T$ and $\eta_F = (E_F - E_c)/k_B T$ have been made. For a device with ideal and symmetrical Ohmic contacts the equilibrium position of the Fermi level in the perovskite is at the conduction band edge meaning $\eta_F = 0$ and $F_{1/2}(0)$. Values for the general function $F_{1/2}(x)$ have been evaluated in [46] giving $F_{1/2}(0)$ = 0.678 (only quoted to 3dp here). Then taking $m^* \approx 0.23 m_e$ in a perovskite conduction band (assumed as MAPbI3 here) [35] where $m_e$ is the rest mass of an electron. We obtain an upper limit of $n_0 \leq 2 \times 10^{18} cm^{-3}$.

Simulation Parameters

| VARIABLE | MAPI | UNIT |
|---|---|---|
| Thickness | $4 \times 10^{-5}$ | cm |
| Electron affinity | -3.8 | eV |
| Ionization potential | -5.4 | eV |
| Equilibrium Fermi energy | -4.6 | eV |
| eDOS conduction band | $1 \times 10^{19}$ | $cm^{-3}$ |
| eDOS valence band | $1 \times 10^{19}$ | $cm^{-3}$ |
| Cation density | $1 \times 10^{18}$ | $cm^{-3}$ |
| Anion density | $1 \times 10^{18}$ | $cm^{-3}$ |
| Electron mobility | 20 | $cm^2 V^{-1} s^{-1}$ |
| Hole mobility | 20 | $cm^2 V^{-1} s^{-1}$ |
| Cation mobility | $1 \times 10^{-10}$ | $cm^2 V^{-1} s^{-1}$ |
| Anion mobility | 0 | $cm^2 V^{-1} s^{-1}$ |
| Relative dielectric constant | 23 | - |
| **Device Boundaries** | | |
| Left hand boundary electron extraction coefficient | $1 \times 10^{10}$ | $cm\ s^{-1}$ |
| Right hand boundary electron extraction coefficient | $1 \times 10^{10}$ | $cm\ s^{-1}$ |
| Left hand boundary hole extraction coefficient | $1 \times 10^{10}$ | $cm\ s^{-1}$ |
| Right hand boundary hole extraction coefficient | $1 \times 10^{10}$ | $cm\ s^{-1}$ |
| $\phi_{m,left}$ | -5.4 | eV |
| $\phi_{m,right}$ | -5.4 | eV |

**Table SI.** Key parameters for metal/perovskite/metal device simulation results presented in figures 1 and 3 in the main text.

| Variable | MAPI | PolyTPD | PTAA | Unit |
|---|---|---|---|---|
| Thickness | $4\times10^{-5}$ | $6\times10^{-7}$ | $2\times10^{-6}$ | cm |
| Electron affinity | -3.8 | -2.3 | -2.3 | eV |
| Ionization potential | -5.4 | -5.2 | -5.3 | eV |
| Equilibrium Fermi energy | -4.6 | -5 | -5.1 | eV |
| eDOS conduction band | $1\times10^{19}$ | $1\times10^{19}$ | $1\times10^{19}$ | cm$^{-3}$ |
| eDOS valence band | $1\times10^{19}$ | $1\times10^{19}$ | $1\times10^{19}$ | cm$^{-3}$ |
| Cation density | $1\times10^{18}$ | - | - | cm$^{-3}$ |
| Anion density | $1\times10^{18}$ | - | - | cm$^{-3}$ |
| Electron mobility | 20 | $7.5\times10^{-3}$ | $7.5\times10^{-3}$ | cm$^2$ V$^{-1}$ s$^{-1}$ |
| Hole mobility | 20 | $7.5\times10^{-3}$ | $7.5\times10^{-3}$ | cm$^2$ V$^{-1}$ s$^{-1}$ |
| Cation mobility | $1\times10^{-10}$ | - | - | cm$^2$ V$^{-1}$ s$^{-1}$ |
| Anion mobility | 0 | - | - | cm$^2$ V$^{-1}$ s$^{-1}$ |
| Relative dielectric constant | 23 | 3 | 3 | - |
| **Device Boundaries** | | | | |
| Left hand boundary electron extraction coefficient | - | - | $1\times10^{10}$ | cm s$^{-1}$ |
| Right hand boundary electron extraction coefficient | - | $1\times10^{10}$ | - | cm s$^{-1}$ |
| Left hand boundary hole extraction coefficient | - | - | $1\times10^{10}$ | cm s$^{-1}$ |
| Right hand boundary hole extraction coefficient | - | $1\times10^{10}$ | - | cm s$^{-1}$ |
| $\phi_{m,left}$ | -5.1 | | | eV |
| $\phi_{m,right}$ | -4.5 | | | eV |

**Table SII.** Key parameters for Au/PTAA/perovskite/PolyTPD/ITO simulations presented in figures 2 and 4 in the main text.